\documentclass[twocolumn,showpacs,superscriptaddress,prl]{revtex4}

\usepackage{latexsym}
\usepackage{indentfirst}
\usepackage{amsmath}
\usepackage{amssymb}
\usepackage{color}
\usepackage{graphicx}
\usepackage{bm}
\usepackage{maybemath}
\newcommand{\be}{\begin{equation}}
\newcommand{\eeq}{\end{equation}}
\newcommand{\bear}{\be\begin{array}}
\newcommand{\bea}{\begin{eqnarray}}
\newcommand{\eea}{\end{eqnarray}}
\newcommand{\nn}{\nonumber}
\newcommand{\dd}{\mathrm{d}}
\newcommand{\ee}{\mathrm{e}}
\newcommand{\ii}{\mathrm{i}}

\begin{document}

\title{Generation of High--Energy Photons with\\
Large Orbital Angular Momentum by Compton Backscattering}

\author{U. D. Jentschura}

\affiliation{Department of Physics, Missouri University of Science and Technology,
Rolla, Missouri 65409, USA}

\affiliation{Institut f\"ur Theoretische Physik,
Universit\"{a}t Heidelberg,
Philosophenweg 16, 69120 Heidelberg, Germany}

\author{V. G. Serbo}

\affiliation{Institut f\"ur Theoretische Physik,
Universit\"{a}t Heidelberg,
Philosophenweg 16, 69120 Heidelberg, Germany}

\affiliation{Novosibirsk State University,
Pirogova 2, 630090, Novosibirsk, Russia}

\begin{abstract}
Usually, photons are described by plane waves with a definite
4-momentum. In addition to plane-wave photons,
``twisted photons'' have recently entered the field of modern laser optics;
these are coherent superpositions of plane waves with a defined
projection $\hbar m$ of the orbital angular momentum onto the propagation axis,
where $m$ is integer. In this paper, we show that it is
possible to produce high-energy twisted photons by Compton
backscattering of twisted laser photons off ultra-relativistic
electrons. Such photons may be of interest for experiments
related to the excitation and disintegration of atoms and nuclei,
and for studying the photo-effect and pair production off nuclei
in previously unexplored experimental regimes.
\end{abstract}


\maketitle

%
%
{\it Introduction.}---An interesting research direction
in modern optics is related to experiments with so-called
``twisted photons.'' These are states of the laser beam whose
photons have a defined value $\hbar m$ of the angular momentum
projection on the beam propagation axis where $m$ is a (large)
integer~\cite{MaVaWeZe2001}. An experimental
realization~\cite{CuKoGr2002} exists for states with projections as
large as $m = 200$. Such photons can be created from usual laser
beams by means of numerically computed holograms. The wavefront of
such states rotates around the propagation axis, and their Poynting
vector looks like a corkscrew (see Fig.~1 in
Ref.~\cite{MaVaWeZe2001}). It was demonstrated that micron-sized
teflon and calcite ``particles'' start to rotate after absorbing twisted
photons~\cite{rotating}.

In this Letter, we show that it is possible to convert twisted
photons from an energy range of about $1\,{\rm{eV}}$ to a higher
energies of up to a hundred GeV using Compton backscattering off
ultra-relativistic electrons.  In principle, Compton
backscattering is an established method for the creation of
high-energy photons and is used successfully in various
application areas from the study of photo-nuclear
reactions~\cite{photonuc,NeTuSh2004} to colliding photon beams of high
energy~\cite{BaBlBl2004}. However, the central question is how to
treat Compton backscattering of twisted photons, whose field
configuration is manifestly different from plane waves.  Below, we
use relativistic Gaussian units with $c=1$, $\hbar = 1$,
$\alpha\approx 1/137$. We denote the electron mass by $m_e$ and
write the scalar product of 4-vectors $k=(\omega,\,{\bm k})$ and
$p=(E,\,{\bm p})$ as $k\cdot p= \omega E - {\bm k} {\bm p}$.

\begin{figure}
\includegraphics[width=1.0\linewidth]{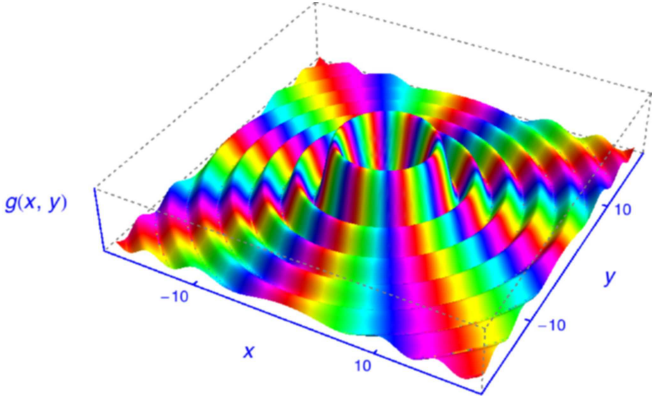}
\caption{\label{fig1} (Color.) The twisted photon vector potential
component ${\cal A}^{\mu}_{\varkappa m k_z \Lambda}(t, x,y,z)$ is
shown for $\mu = 2$ ($y$ component), $\varkappa = 1$, $m = 5$,
$k_z = \sqrt{24}$, and $\Lambda = 1$. The plot displays $g(x,y) =
| {\cal A}^{\mu}_{\varkappa m k_z \Lambda}(0, x,y,0)|^2$ as a
function of $x$ and $y$. The complex phase of the vector
potential, which is a superposition of terms proportional to
$\ee^{4 \ii \varphi_r}$ and $\ee^{6 \ii \varphi_r}$, is indicated
by the variation of the color of the wave function on the rainbow
scale.}
\end{figure}

%
%
{\it Twisted photon.}---We wish to construct a twisted
photon state with definite longitudinal
momentum~$k_z$, absolute value of transverse momentum $\varkappa$
and projection $m$ of the orbital angular momentum onto the $z$
axis (propagation axis).
We start from a
plane-wave photon state with 4-momentum $k=(\omega, {\bm k})$ and
helicity $\Lambda=\pm 1$,
\begin{align}
\label{photon} A^\mu_{k\Lambda}(t, \bm r) =& \;\sqrt{4\pi}\,
e^\mu_{k\Lambda}\, \frac{\ee^{-\ii (\omega \, t-{\bm k}  {\bm
r})}}{\sqrt{2\omega}}\,,
\end{align}
where $e^\mu_{k\Lambda}$ is the polarization 4-vector of the
photon
($e_{k\Lambda} \cdot k = 0$ and
$e^*_{k\Lambda} \cdot e_{k\Lambda'} = -\delta_{\Lambda \Lambda'}$,
with $\Lambda = -1,1$).
The twisted photon vector potential
${\cal A}^\mu_{\varkappa m k_z \Lambda}(r,\varphi_r,z,t)$ is obtained
after integration over the conical transverse momentum
components ${\bm k}_\perp = (k_x, k_y, 0)$ of the wave vector
$\bm k = (k_x,\,k_y,\, k_z)$, with amplitude
\begin{equation}
a_{\varkappa m}({\bm k}_\perp) = (-\ii)^m \; \ee^{\ii m\varphi_k}
\; \sqrt{\frac{2\pi}{\varkappa}}\, \delta(k_\perp-\varkappa) \,.
\end{equation}
Here, $\varphi_k$ is the azimuth angle of ${\bm k}_\perp$, and
\begin{align}
\label{twistedwave}
& {\cal A}^\mu_{\varkappa m k_z \Lambda}(r,\varphi_r,z,t) =
\int a_{\varkappa m}({\bm k}_\perp) \,
A^\mu_{k\Lambda}(t, \bm r) \, \frac{\dd^2 k_\perp}{(2\pi)^2}
\nonumber\\
& =
\sqrt{\frac{4\pi}{2\omega}} \; \ee^{-\ii (\omega t - k_z z)}\;
\int e_{k \Lambda} \, a_{\varkappa m}({\bm k}_\perp)\,
\ee^{\ii{\bm k}_\perp {\bm r}}
\frac{\dd^2k_\perp}{(2\pi)^2} \,.
\end{align}
Furthermore, $\omega=|{\bm k}|= \sqrt{\varkappa^2+k_z^2}$.
For further analysis we introduce the three
four-vectors
\begin{equation}
\eta^{(\pm)}=\mp \tfrac{1}{\sqrt{2}}\,(0,\,1,\pm \ii,\,0)\,,\;\;
\eta^{(z)}=(0,\,0,\,0,\,1) \,.
\label{vec}
\end{equation}
The initial twisted photon is composed of wave vectors
of the form
\begin{equation}
\label{kin}
{\bm k} = \omega \; (\sin{\alpha_0} \, \cos{\varphi_k}, \;
\sin{\alpha_0} \, \sin{\varphi_k}, \; -\cos{\alpha_0}) \,,
\end{equation}
which for $\alpha_0 = 0$ propagate in the negative $z$ direction.
Here, $\theta=\pi-\alpha_0$ and $\varphi_k$ are the polar and
azimuth angles of the initial photon, and we have
$\tan{\alpha_0} = k_\perp / (-k_z)$. The polarization vectors
can be expressed as
\begin{align}
\label{e1}
e_{k \Lambda}=& \;
\eta^{(-\Lambda)} \, \ee^{\ii \Lambda \varphi_k}\,
\cos^2\left(\frac{\alpha_0}{2}\right) +
\eta^{(\Lambda)} \, \ee^{-\ii\Lambda \varphi_k}\,
\sin^2\left(\frac{\alpha_0}{2}\right)
\nn\\
& \; + \frac{\Lambda}{\sqrt{2}}\,\eta^{(z)}\,
\sin{\alpha_0}\,.
\end{align}
Integration leads to
\begin{align}
\label{psi}
& \int e_{k \Lambda} \, a_{\varkappa m}({\bm k}_\perp)\,
\ee^{\ii{\bm k}_\perp {\bm r}}
\frac{\dd^2k_\perp}{(2\pi)^2}
= \frac{\Lambda}{\sqrt{2}}\, \eta^{(z)}\, \psi_{\varkappa m}(r,\varphi_r)\,
\sin{\alpha_0}
\nonumber \\
& \qquad + \ii^{\Lambda} \eta^{(-\Lambda)}
\psi_{\varkappa, m+\Lambda}(r,\varphi_r)
\cos^2\left( \frac{\alpha_0}{2}\right)
\nonumber \\
& \qquad - \ii^{\Lambda} \eta^{(\Lambda)}
\psi_{\varkappa, m-\Lambda}(r,\varphi_r)
\sin^2 \left( \frac{\alpha_0}{2} \right) \,,
\end{align}
with the scalar twisted particle wave function
\begin{equation}
\psi_{\varkappa m}(r,\varphi_r) =
\frac{\ee^{\ii m\varphi_r}}{\sqrt{2\pi}}\,
\sqrt{\varkappa}\, J_m(\varkappa \; r)\,.
\end{equation}
The vector field ${\cal A}^\mu_{\varkappa m k_z
\Lambda}(r,\varphi_r,z,t)$ describes a photon state with
projections of the orbital angular momentum on the $z$ axis equal
to $m-1,\,m,\,m+1$. For large $m$, the restriction to
$(m-1,m,m+1)$ means that the twisted state is a state with
``almost defined angular momentum projection $m$''
(see Fig.~\ref{fig1}), and we denote it as $|\varkappa, m,
k_z, \Lambda\rangle$.

The usual $S$ matrix element for plane-wave (PW) Compton
scattering involves an electron being scattered from the state
$|p, \lambda \rangle$ with 4-momentum $p$ and helicity
$\lambda=\pm \tfrac12$ to the state $|p', \lambda' \rangle$ and a
photon being scattered from the state $|k, \Lambda\rangle$ to the
state $|k', \Lambda' \rangle$,
\begin{equation}
\label{Sfi}
S^{\rm (PW)}_{fi} \equiv \langle k',\Lambda', p',\lambda'| S
|k, \Lambda, p,\lambda\rangle \,.
\end{equation}
In view of Eq.~\eqref{twistedwave}, the $S$ matrix element $S^{\rm
(TW)}_{fi}$ for the scattering of a twisted (TW) photon $|\varkappa,
m,k_z, \Lambda\rangle$ into the state $|\varkappa', m', k'_z,
\Lambda' \rangle$ needs to be integrated as follows,
\begin{align}
\label{convol} & S^{\rm (TW)}_{fi} \equiv
\langle \varkappa', m',k'_z, \Lambda'; p', \lambda' | S |
\varkappa, m, k_z, \Lambda; p, \lambda \rangle
\nonumber\\
& = \int \frac{\dd^2k_\perp}{(2\pi)^2}\,\frac{\dd^2k'_\perp}{(2\pi)^2} \,
a^*_{\varkappa' m'}({\bm k}'_\perp)
\; S^{\rm (PW)}_{fi} \; a_{\varkappa m}({\bm k}_\perp) \,.
\end{align}
%

%
%
{\it Compton scattering of plane-wave photons.}---We investigate
the collision of an ultra-relativistic electron with 4-momentum
$p=(E,0,0,v \; E)$, $v= |{\bm p}|/E$, and $\gamma=E/m_e$,
propagating in the positive $z$ direction, and a photon of energy
$\omega$ and three-momentum given by Eq.~\eqref{kin}.
After the scattering, the
4-momentum of the electron is $p'$, and the scattered photon has
energy $\omega'$ and three-momentum
${\bm k}'=\omega' \; (\sin{\theta'} \cos{\varphi'_k}, \;
\sin{\theta'} \sin{\varphi'_k}, \; \cos{\theta'})$,
where $\theta'$ and $\varphi'_k$ are the polar and azimuth angles
of the final photon.
From the equation $(p+k-k')^2=m_e^2$, we obtain
\begin{equation}
\label{omf}
\omega' =
\frac{m_e^2 \, x}{2E(1-v\cos{\theta'}) +
2\omega(1+\cos{\beta})}\,,
\end{equation}
where $x \, m_e^2 = 2 \, p \cdot k = 2\omega E(1+v\cos{\alpha_0})$
and $\cos{\beta}= \cos{\alpha_0} \, \cos{\theta'} -\sin{\alpha_0}
\, \sin{\theta'} \, \cos{(\varphi_k-\varphi'_k})$.
The $S$-matrix element for plane waves is
\begin{equation}
\label{Sfidef}
S^{\rm (PW)}_{fi}=\ii\,(2\pi)^4 \,\delta(p+k-p'-k')\;
\frac{M_{fi}}{4\sqrt{E\,E'\,\omega\,\omega'}}\,,
\end{equation}
where the amplitude $M_{fi}$ in the Feynman gauge is
\begin{subequations}
\begin{eqnarray}
\label{defM}
M_{fi} &=&
4\pi\alpha\left(\frac{A}{s-m_e^2}+\frac{B}{u-m_e^2}\right)\,,\\
\label{defA}
A &=& \bar{u}_{p'\lambda'} \,
\hat{e}^*_{k' \Lambda'} \,
\left(\hat p +\hat k +m_e\right)
\hat{e}_{k \Lambda}\, u_{p \lambda}\,,
\\
\label{defB} B &=& \bar{u}_{p' \lambda'} \, \hat{e}_{k \Lambda}
\left(\hat{p}' -\hat k +m_e\right) \hat{e}^*_{k' \Lambda'} \, u_{p
\lambda} \,,
\end{eqnarray}
\end{subequations}
and $s-m_e^2=x \, m^2_e$, $m_e^2-u= 2 \, p \cdot k'= 2 \, \omega'
E \, (1-v\cos{\theta'})$. The bispinors $u_{p \lambda}$ and $u_{p'
\lambda'}$ describe the initial and final electrons, and $e_{k
\Lambda}$ and $e_{k' \Lambda'}$ are the polarization vectors of
the initial and final photon. We denote the Feynman dagger as
$\hat p = \gamma^\mu p_\mu$.
Using Dirac algebra, we may write $A = A_1 + A_2$ and
$B=B_1+B_2$ with
\begin{subequations}
\begin{align}
\label{A1}
A_1 =& \;  -\bar{u}_{p'\lambda'} \;
\hat{e}_{k'\Lambda'}^* \;
\hat{e}_{k\Lambda} \;
\hat{k}\, u_{p\lambda}\,,
\\
\label{A2}
A_2 =& \; 2 (e_{k\Lambda} \cdot p)\;
\bar{u}_{p'\lambda'} \;
\hat{e}_{k'\Lambda'}^*\; u_{p\lambda}\,,
\\
\label{B1}
B_1 =& \;  \bar{u}_{p'\lambda'} \;
\hat{k} \;
\hat{e}_{k\Lambda} \;
\hat{e}_{k' \Lambda'}^* \;
u_{p \lambda}\,,\;\;
\\
\label{B2}
B_2 =& \; 2 (e_{k\Lambda} \cdot p') \;
\bar{u}_{p' \lambda'} \;
\hat{e}_{k'\Lambda'}^*\; u_{p\lambda}\,.
\end{align}
\end{subequations}
$M_{fi}$ as defined in
Eq.~\eqref{defM} can thus be written as
\begin{subequations}
\label{Mfi12}
\begin{align}
M_{fi}=& \; M^{(1)}_{fi}+M^{(2)}_{fi} \,,
\\
M^{(1,2)}_{fi} =& \;
4\pi\alpha\left(\frac{A_{1,2}}{s-m_e^2}+
\frac{B_{1,2}}{u-m_e^2}\right)\,.
\end{align}
\end{subequations}
For a head-on collision of a plane-wave photon and electron, the
relativistic kinematics then imply the following differential
cross section, for the unpolarized cross section (summation over
the outgoing and averaging over the incoming electron and photon
polarizations),
\begin{align}
\label{cs}
\frac{\dd\sigma}{\dd\Omega'} =&\;
\frac{2\alpha^2\gamma^2}{m_e^2}\,F(x,n) \,,
\quad
n \equiv \gamma \, \theta' \,,
\quad
x = \frac{4\omega E}{m_e^2} \,,
\nonumber\\
F(x,n) =& \; \left(\frac{1}{1+x+n^2}\right)^2
\left[\frac{1+x+n^2}{1+n^2}
+ \frac{1+n^2}{1+x+n^2} \right.
\nonumber\\
& \; \left.  -4\frac{n^2}{(1+n^2)^2} +
{\mathcal O}\left(\gamma^{-1} \right)\right] \,.
\end{align}
According to Eq.~\eqref{twistedwave}, a twisted photon is a
superposition of plane-wave photons with the same energy and
conical momentum spread. We thus expect that the twisted and
plane-wave photon scattering cross section will be related.
Indeed, for the mixed (m) case where the initial photon is twisted
but the outgoing one is a plane-wave photon, one finds
\begin{align}
\label{mixed}
S^{\rm (m)}_{fi} \equiv & \;
\langle k', \Lambda', p', \lambda' | S |
\varkappa, m, k_z, \Lambda; p, \lambda \rangle
\nonumber\\
=& \; \int \frac{\dd^2k_\perp}{(2\pi)^2}\,
S^{\rm (PW)}_{fi} \; a_{\varkappa m}({\bm k}_\perp) \,,
\end{align}
and the corresponding cross section is given by
Eq.~\eqref{cs} with the only replacement
\begin{equation}
\label{replace}
x = \frac{4\omega E}{m_e^2} \to
\frac{4\omega E \; \cos^2\alpha_0}{m_e^2} \,.
\end{equation}
For strict backscattering geometry, the differential Compton cross
section~\eqref{cs} and the energy of the scattered photon attain
maxima, and additional simplifications are possible because the
azimuth angle of the photon $\varphi'_k = \varphi_k$ is conserved,
as discussed in the following.

\begin{figure}
\includegraphics[width=0.9\linewidth]{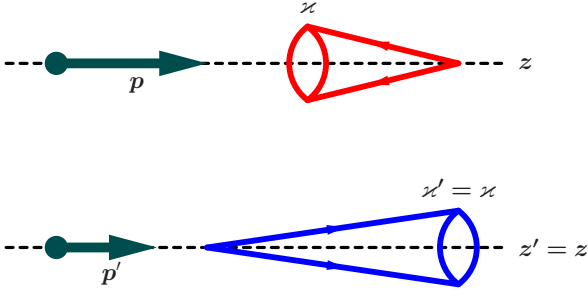}
\caption{\label{fig2} (Color.) Initial (above) and final
(below) states for the head-on Compton backscattering geometry of
a twisted photon. According to Eq.~\eqref{mainres}, the conical
momentum spread $\varkappa$ of the initial twisted photon is
preserved ($\varkappa' = \varkappa$) during the scattering, but
the propagation  energy increases: $\omega' = \sqrt{{k'}_z^2 +
{\varkappa'}^2} \gg \omega = \sqrt{k_z^2 + \varkappa^2}$.}
\end{figure}

%
%
{\it Compton backscattering of twisted photons.}---For twisted photons, the final
$m'$ photon is a superposition of plane waves with
small transverse momentum ${\bm k}'_\perp={\bm k}_\perp$ and very
small scattering angle $ \theta'=k'_\perp/\omega'\lesssim
(1+x)/(4\gamma^2)$ (see Fig.~\ref{fig2}).
In this limit, $\omega'= x \, E/(1+x)$. For
strict backward scattering, several quantum numbers
in Eq.~(\ref{convol}) are thus conserved under the
scattering for twisted (TW) photons,
\begin{align}
\label{tildeSfi}
& S^{\rm (TW)}_{fi} = 2\pi\,\ii^{m'-m+1}
\delta(\varkappa-\varkappa') \delta(E+\omega-E'-\omega') \\
& \times \delta(p_z+k_z-p_z'-k_z')
\int\limits_0^{2\pi}{\ee^{\ii(m-m')\varphi_k}}\,
\frac{M^{(1)}_{fi} + M^{(2)}_{fi}}{4\sqrt{EE'\omega\omega'}}\,
\dd\varphi_k \,,
\nonumber
\end{align}
where we have used the decomposition~\eqref{Mfi12}.

In order to carry out the integration over $\varphi_k$, we have to
analyze the dependence of the polarization vectors $e_{k \Lambda}$
and $e_{k'  \Lambda'}$ on the azimuth angle. To this end, we write
the polarization vector of the final photon in the scattering
amplitude $M_{fi}$ in the form $e_{k'  \Lambda'} =
-\Lambda'\,(e^{(x')}+\ii \Lambda' e^{(y')})/\sqrt{2}$, where the
unit vector $e^{(x')}=(0,\,{\bm e}^{(x')})$ is in the scattering
plane, defined by the vectors $\bm p
\parallel \bm p'$ and $\bm k'$, while the unit vector $e^{(y')}$
is orthogonal to it: ${\bm e}^{(x')}\, \parallel \; ({\bm p}\times
{\bm k}')\times {\bm k}'$ and ${\bm e}^{(y')}\,
\parallel \; ({\bm p}\times {\bm k}')$. As a result, we have in
4-vector component notation
\begin{equation}
\label{e2} e_{k' \Lambda'} = -\frac{\Lambda'}{\sqrt{2}}\,
\left( \begin{array}{c} 0 \\
\cos{\theta'} \cos{\varphi_k}-\ii\Lambda'\sin{\varphi_k}\\
\cos{\theta'} \sin{\varphi_k}+\ii\Lambda'\cos{\varphi_k}\\
-\sin{\theta'}
\end{array} \right) \,.
\end{equation}
Omitting terms of the order of $\theta'$, we obtain
\begin{equation}
\label{e2app} e_{k'  \Lambda'} = -
\frac{\Lambda'}{\sqrt{2}}\,(0,\,1,\,\ii\Lambda',\,0)\,
\ee^{-\ii\Lambda'\varphi_k} = \eta^{(\Lambda')}\,
\ee^{-\ii\Lambda'\varphi_k}\,.
\end{equation}
The polarization vector $e_{k  \Lambda}$ of a ``conical''
component of the initial twisted photon (as a function of
$\varphi_k$) is obtained by setting $\theta'=\pi-\alpha_0$ in
$e_{k'  \Lambda'}$ and coincides with Eq.~\eqref{e1}.
Substituting the expressions for $e_{k' \Lambda'}$ and $e_{k
\Lambda}$ given in Eqs.~\eqref{e2app} and~\eqref{e1} into the
definitions of $A_1$ and $B_1$ given in Eqs.~\eqref{A1}
and~\eqref{B1}, we find for twisted photons,
\begin{subequations}
\label{A1B1}
\begin{align}
 \label{A1s}
A_1 = & \; 2 \omega \, \sqrt{EE'}\, \left[(1-\Lambda \,
\Lambda'\cos{\alpha_0}) \, (1+\cos{\alpha_0}) \right.
\nonumber\\
& \; \left. +2\lambda \,\Lambda \, \sin^2{\alpha_0}
\right]\,\delta_{\lambda\lambda'}\,\delta_{2\lambda, -\Lambda'}\,, \\
 \label{B1s}
B_1 = & \; -2 \omega \, \sqrt{EE'}\, \left[(1-\Lambda \,
\Lambda'\cos{\alpha_0}) \, (1+\cos{\alpha_0}) \right.
\nonumber\\
& \; \left. -2\lambda \,\Lambda \, \sin^2{\alpha_0}
\right]\,\delta_{\lambda\lambda'}\,\delta_{2\lambda,\Lambda'}\,.
\end{align}
\end{subequations}
One may write the neglected contribution  $M^{(2)}_{fi}$ as
$M^{(2)}_{fi} = -4\pi\alpha\; \bar{u}_{p'\lambda'} \,
\hat{e}_{k'\Lambda'}^*\,u_{p\lambda} \, (e_{k\Lambda})_z \,
\epsilon / \omega$. It is negligible for our relativistic
kinematics ($\gamma \gg 1$), because
\begin{align}
\epsilon =& \; \omega\,\left(\frac{p_z}{k \cdot p} - \frac{p'_z}{k
\cdot p'}\right) = \frac{x(x+2)}{2\gamma^2(1+\cos{\alpha_0})^2}
\ll 1 \,.
\end{align}
Therefore, we have $|M^{(2)}_{fi}| \ll |M^{(1)}_{fi}|$ for strict backward
scattering, and the $S$ matrix element reads
\begin{align}
\label{mainres}
S^{\rm (TW)}_{fi} \approx & \; \ii \, (2 \pi)^2 \, \delta_{m m'} \,
\delta(\varkappa-\varkappa') \,\delta(E+\omega-E'-\omega')
\nonumber\\
& \; \times \; \delta(p_z+k_z-p_z'-k_z')
\frac{M^{(1)}_{fi}}{4\sqrt{EE'\omega\omega'}} \,,
\end{align}
with $M^{(1)}_{fi}$ given in Eqs.~\eqref{Mfi12} and \eqref{A1B1}.
This result states that for strict backscattering, the angular
momentum projection $m'= m$ and the conical momentum spread
$\varkappa' = \varkappa$ of the twisted photons are conserved and
confirms the principal possiblity for the frequency upconversion
of twisted photons under strict Compton backscattering. A
technique for the registration of electrons scattered at small
(even zero) angles after the loss of energy in the Compton process
is implemented, for example, in the device for backscattered
Compton photons installed on the VEPP-4M collider
(Novosibirsk)~\cite{NeTuSh2004}.

Certainly, it is interesting to estimate the admixture of
different $m' \neq m$ twisted photon states if the electron
carries away a small transverse momentum $p'_\perp \ll \varkappa$
under the scattering.
A solution of the relativistic kinematic equations then implies
that the azimuthal angle of the scattered twisted photon
component is not conserved but acquires a phase slip,
\begin{equation}
\varphi_{k'} = \varphi_{k} + \Delta \,, \quad \Delta \approx
\frac{p'_\perp}{\varkappa}\,\sin\varphi_{k} \,.
\end{equation}
Taking into account this phase slip we obtain  a distribution
approximately given by Eq.~\eqref{mainres} under the replacement
\begin{align}
&\int_0^{2 \pi} \frac{\dd \varphi_k}{2 \pi} \, \ee^{\ii \,
(m\varphi_{k}-m'\varphi_{k'})}\big|_{\varphi_{k'}=\varphi_{k}}=
\delta_{m m'}
 \\
&  \to  \int_0^{2 \pi} \frac{\dd \varphi_k}{2 \pi} \, \ee^{\ii \,
[(m-m')\varphi_{k}-m'\Delta]}= J_{m-m'}\left(
m'p'_\perp/\varkappa\right)\,. \nonumber
\end{align}
This yields a distribution where the scattered twisted photon
angular momenta $m'$ are displaced from the initial twisted photon
angular momentum $m$ by $\delta m \equiv |m' - m| \sim m' \,
p'_\perp/\varkappa$.  Finally, as the initial twisted photon is
obtained by an integration over a conical angular distribution of
plane-wave components,  the energy of the final twisted photon for
the case of non-strict backscattering can be obtained from
Eq.~\eqref{omf}.

%
%
{\it Conclusions.}---The general convoluted invariant
matrix element~\eqref{convol} for Compton
scattering of twisted photons, which can be
evaluated numerically for arbitrary scattering geometry, is found to take a
particularly simple form for strict backscattering (see Fig.~\ref{fig2}),
according to Eqs.~\eqref{Mfi12} and~\eqref{A1B1}. For that geometry, the energy of
the final twisted photon is increased most effectively ($\omega'/\omega \sim
\gamma^2 \gg 1$). According to Eq.~\eqref{mainres}, the magnetic quantum number
$m'=m$ and  the conical momentum spread $\varkappa' = \varkappa$ are preserved
under strict backscattering.
This implies that the conical angle $\theta'$ of the scattered twisted photon
is very small, $\theta' \approx \varkappa'/\omega' \sim 1/\gamma^2$.

High-energy photons with large orbital angular momenta projections can be used
for experimental studies regarding the excitation of atoms into circular
Rydberg states, and for studying the photo-effect and the ionization of atoms,
as well as the pair production off nuclei. As ion traps for highly-charge ions
are currently under construction (e.g., Ref.~\cite{HITRAP}),
one of the most interesting experiments would concern the question of whether
nuclear fission can be achieved by the absorption of fast rotating nuclei, via
the absorption of one or more twisted high-$m$ photons
at energies below the giant
dipole resonances~\cite{HaWo2001} which are typically
in the range of $\sim$~10--30~MeV.
Such a study might reveal fundamental insight into the dynamics of a fast
rotating quantum many-body system.

The authors are grateful to I.~Ginzburg, D.~Ivanov, I.~Ivanov, G.~Kotkin, N.~Muchnoi,
O.~Nachtmann, V.~Telnov, A.~Voitkiv, V.~Zelevinsky and V.~Zhilich for useful
discussions. This research was supported by the National Science Foundation
(PHY-8555454) and by the Missouri Research Board. V.G.S.~is supported by the
Russian Foundation for Basic Research via grants 09-02-00263 and
NSh-3810.2010.2.

\end{document}